\documentclass[a4paper]{article}

\usepackage{16lomcon}        % Proceedings volume layout metrics
\usepackage{cite}             % Smart range citations
\usepackage{epsfig}           % Encapsulated PostScript figure inclusion

\begin{document}

%%%%%%%%%%%%%%%%%%%%%%%%%%%%%%%%%%%%%%%%%%%%%%%%%%%%%%%%%%%%%%%%%%
% The preamble of the paper
%%%%%%%%%%%%%%%%%%%%%%%%%%%%%%%%%%%%%%%%%%%%%%%%%%%%%%%%%%%%%%%%%%

\title{REVIEW OF DOUBLE BETA DECAY EXPERIMENTS}

\author{Alexander Barabash \email{barabash@itep.ru}
%        and
%        Ivan Author \email{ivan.author@address.com}
}

\affiliation{Institute of Theoretical and Experimental Physics, 
B. Cheremushkinskaya 25, 117218
Moscow, Russia}

% You may repeat \author and \affiliation as many times as necessary!

\date{}
% Print it out!
\maketitle

%%%%%%%%%%%%%%%%%%%%%%%%%%%%%%%%%%%%%%%%%%%%%%%%%%%%%%%%%%%%%%%%%%
% The preamble of the paper
%%%%%%%%%%%%%%%%%%%%%%%%%%%%%%%%%%%%%%%%%%%%%%%%%%%%%%%%%%%%%%%%%%

\begin{abstract}
The brief review of current experiments on search and studying of double 
beta decay processes is done. Best present limits on $\langle m_{\nu} \rangle$
and $\langle g_{ee} \rangle$ are presented. 
%Results of the most sensitive 
%experiments (EXO, KamLAND-Zen and GERDA) are discussed and values 
%of modern limits on effective Majorana neutrino mass ($\langle m_{\nu}\rangle$) are given. 
%New results on two neutrino double beta decay are presented.
%     In the second part of the review 
Prospects of search for 
neutrinoless double beta decay in new experiments with sensitivity 
to $\langle m_{\nu}\rangle$ at the level of $\sim$ 0.01-0.1 eV are discussed. 
%Parameters 
%and characteristics of the most perspective projects (CUORE, 
%GERDA, MAJORANA, SuperNEMO, EXO, KamLAND-Zen and SNO+) are given.

\end{abstract}

\section{Introduction}

Interest in $0\nu\beta\beta$ decay has seen a significant renewal in recent 
10 years after evidence for neutrino oscillations was obtained 
from the results of atmospheric, solar, reactor, and accelerator 
neutrino experiments. These results are impressive proof that 
neutrinos have a nonzero mass. The detection and study of $0\nu\beta\beta$ 
decay may clarify the following problems of neutrino physics: 
(i) lepton number non-conservation, (ii) neutrino nature: whether 
the neutrino is a Dirac or a Majorana particle, (iii) absolute 
neutrino mass scale (a measurement or a limit on m$_1$), (iv) the 
type of neutrino mass hierarchy (normal, inverted, or quasidegenerate), 
(v) CP violation in the lepton sector (measurement of the Majorana 
CP-violating phases).

%Progress in the double beta decay is connected with increase in mass 
%of a studied isotope and sharp decrease in a background.
%In 2011 the EXO--200 \cite{ACK11} and KamLAND--Zen \cite{GAN12} installations in which 
%hundreds kilograms of $^{136}$Xe are used and GERDA-I experiment with $\sim$ 15 kg of $^{76}$Ge
%were started. Very low level of background was reached in all of these experiments. Soon 
%it is planned to carry out start of several more installations with 
%ass of studied isotopes $\sim 100$ kg (SNO+ \cite{HAR12} and CUORE \cite{GOR12}). And it 
%means the beginning of a new era in $2\beta$ decay experiments when 
%sensitivity to effective Majorana mass of neutrino will reach for 
%the first time values $< 0.1$ eV.

%Let us consider three main modes of $\beta\beta$ decay:

%\begin{equation}
%(A,Z) \rightarrow (A,Z+2) + 2e^{-} + 2\bar { \nu},
%\end{equation}

%\begin{equation}
%(A,Z) \rightarrow (A,Z+2) + 2e^{-},
%\end{equation}

%\begin{equation}
%(A,Z) \rightarrow (A,Z+2) + 2e^{-} + \chi^{0}(+ \chi^{0}).
%\end{equation}

%This review is structured as follows. In Section 2 I summarize the main 
%achievements in search for double beta decay by present time. Section 3 is 
%dedicated to description of current large-scale experiments. The short review 
%of most developed future experiments are presented in Section 3. 
%My conclusions are summarizes in Section 4.

\section{Present status}

\subsection{Two neutrino double beta decay}%\label{subsec:general}

The $2\nu\beta\beta$ decay ($(A,Z) \rightarrow (A,Z+2) + 2e^{-} + 2\bar { \nu}$) is a 
second-order process, which is not forbidden 
by any conservation law. The detection of this process provides the experimental 
determination of the nuclear matrix elements (NME) involved in the $\beta\beta$-decay processes. 
This leads to the development of theoretical schemes for NME calculations both in connection with the
$2\nu\beta\beta$ decays as well as the $0\nu\beta\beta$ decays. Table 1 displays the present-day 
averaged and recommended values of $T_{1/2}(2\nu)$ from \cite{BAR13} 

\begin{table}[ht]
\label{Table1}
\caption{Average and recommended $T_{1/2}(2\nu)$ values (from
\cite{BAR13}).}
\vspace{0.5cm}
%\rule[-2mm]{0mm}{5mm}
\begin{center}
\begin{tabular}{cc}
\hline
Isotope & $T_{1/2}(2\nu)$, yr \\
\hline
$^{48}$Ca & $4.4^{+0.6}_{-0.5}\cdot10^{19}$ \\
$^{76}$Ge & $1.6^{+0.13}_{-0.10} \cdot10^{21}$ \\
$^{82}$Se & $(0.92 \pm 0.07)\cdot10^{20}$ \\
$^{96}$Zr & $(2.3 \pm 0.2)\cdot10^{19}$ \\
$^{100}$Mo & $(7.1 \pm 0.4)\cdot10^{18}$ \\
$^{100}$Mo-$^{100}$Ru$(0^{+}_{1})$ & $6.2^{+0.7}_{-0.5}\cdot10^{20}$ \\
$^{116}$Cd & $(2.85 \pm 0.15)\cdot10^{19}$\\
$^{128}$Te & $(2.0 \pm 0.3)\cdot10^{24}$ \\
$^{130}$Te & $(6.9 \pm 1.3)\cdot10^{20}$ \\
$^{136}$Xe & $(2.20 \pm 0.06)\cdot10^{21}$ \\
$^{150}$Nd & $(8.2 \pm 0.9)\cdot10^{18}$ \\
$^{150}$Nd-$^{150}$Sm$(0^{+}_{1})$ & $1.33^{+0.45}_{-0.26}\cdot10^{20}$\\
$^{238}$U & $(2.0 \pm 0.6)\cdot10^{21}$  \\
$^{130}$Ba; ECEC(2$\nu$) & $\sim 10^{21}$  \\
\hline
\end{tabular}
\end{center}
\end{table}

\subsection{Neutrinoless double beta decay}%\label{subsec:general}

The $0\nu\beta\beta$ decay ($(A,Z) \rightarrow (A,Z+2) + 2e^{-}$) violates the law of lepton-number 
conservation ($\Delta$L = 2) 
and requires that the Majorana neutrino has a nonzero rest mass or that an admixture of 
right-handed currents be present in weak interaction. Also, this process is possible 
in some supersymmetric models, where $0\nu\beta\beta$ decay is initiated by the exchange 
of supersymmetric 
particles. This decay also arises in models featuring an extended Higgs sector within 
electroweak-interaction theory
and in some other cases (see review \cite{VER12}, for example). The best present limits
on $0\nu\beta\beta$ decay are presented in Table 2. In calculating constraints on 
$\langle m_{\nu} \rangle$, the NMEs from \cite{SUH12,SIM13,BAR09,RAT10,ROD10,MEN09}, 
phase-space factors from \cite{KOT12} and $g_A$ = 1.27 were used.
Present conservative limit on $\langle m_{\nu} \rangle$ can be set as 0.34 eV. 

\begin{table}[ht]
\label{Table2}
\caption{Best present results on $0\nu\beta\beta$ decay (limits at
90\% C.L.).}
\vspace{0.5cm}
%\rule[-2mm]{0mm}{5mm}
\begin{center}
\begin{tabular}{ccccc}
\hline
Isotope & $T_{1/2}$, y & $\langle m_{\nu} \rangle$, eV; 
\cite{SUH12,SIM13,BAR09,RAT10,ROD10,MEN09} & Experiment \\
%& & \cite{SUH12,SIM13,BAR09,RAT10,ROD10,MEN09}   \\
\hline
$^{76}$Ge & $>2.1\cdot10^{25}$ & $<0.25-0.62$ & GERDA-I \cite{AGO13} \\
%& $\simeq 1.2\cdot10^{25}$(?) & $\simeq 0.5-1.3(?)$ & $\simeq
%0.7(?)$ & Part of HM \cite{KLA04} \\
%& $>1.6\cdot10^{25}$ & $<0.36-0.92$ & $<0.58-0.64$ & IGEX
%\cite{AAL02} \\
%\hline
$^{100}$Mo & $>1.1\cdot10^{24}$ & $<0.34-0.87$ & NEMO-
3 \cite{ARN13} \\
$^{130}$Te & $>2.8\cdot10^{24}$ & $<0.31-0.76$ & CUORICINO \cite{AND11} \\
$^{136}$Xe & $>1.9\cdot10^{25}$ & $<0.14-0.34$ & KamLAND-Zen
\cite{GAN13} \\
%$^{116}$Cd & $>1.3\cdot10^{23}$ & $<1.4-2.5$ & $<3.7-4.3$ &
%SOLOTVINO \cite{DAN03} \\
%$^{82}$Se & $>1\cdot10^{23}$ & $<1.7-3.7$ & $<3.8-4.7$ & NEMO-3
%\cite{ARN05} \\
\hline
\end{tabular}
\end{center}
\end{table}

\subsection{Neutrinoless double beta decay with Majoron emission}%\label{subsec:general}

The $0\nu\chi^0\beta\beta$ decay($(A,Z) \rightarrow (A,Z+2) + 2e^{-} + \chi^{0}$) 
requires the existence of a Majoron. It is a massless 
Goldstone boson that arises due to a global breakdown of (B-L) symmetry, where B and L 
are, respectively, the baryon and the lepton number. The Majoron, if it exists, could playa 
significant role in the history of the early Universe and in the evolution of stars.
The best obtained limits
on $0\nu\chi^0\beta\beta$ decay are presented in Table 3 \footnote{I would like
 to stress that I use here phase-space factor values (G) from \cite{DOI88}. 
These values are $\sim$ 2 times higher then in \cite{SUH98}, for example. 
This difference can be explain as follows: in \cite{SUH98} "canonical" 
formulas from \cite{DOI85} to calculate 
G were used. But in \cite{DOI88} 
it was mentioned that factor 2 was lost in \cite{DOI85}. }. 
Present conservative limit can be set as 
$\langle g_{ee} \rangle < 10^{-5}$.

\begin{table}[ht]
\label{Table3}
\caption{Best present results on $0\nu\chi^{0}\beta\beta$ decay
(ordinary Majoron) at 90\% C.L. The NME from \cite{SUH12,SIM13,BAR09,RAT10,ROD10,MEN09}, 
phase-space factors from \cite{DOI88} and $g_A$ = 1.27 were used.}
\vspace{0.5cm}
%\rule[-2mm]{0mm}{5mm}
\begin{center}
\begin{tabular}{cccc}
\hline
Isotope & $T_{1/2}$, y & $\langle g_{ee} \rangle$, $\times 10^{-5}$
 & Experiment \\  
\hline
$^{76}$Ge & $>6.4\cdot10^{22}$  & $< 6.3-15.8 $ & Heidelberg-Moscow \cite{KLA01} \\
$^{82}$Se & $>1.5\cdot10^{22}$  & $< 5.0-11.6 $ & NEMO-3 \cite{ARN06}\\
$^{100}$Mo & $>4.4\cdot10^{22}$ & $< 1.6-4.1 $ & NEMO-3 \cite{ARN13} \\
$^{136}$Xe & $> 2.6\cdot10^{24}$ & $< 0.4-1.0 $ & KamLAND-Zen \cite{GAN12} \\
%$^{128}$Te & $>2\cdot10^{24}$(geochem)\cite{MAN91} & $<(0.7-
%1.6)\cdot10^{-4}$ & $<(1.9-2.4)\cdot10^{-4}$ \\
\hline
\end{tabular}
\end{center}
\end{table}

\section{Current large-scale experiments}

%In this section the current large-scale experiments are discussed.

\subsection{EXO-200}

EXO--200 (Enriched Xenon Observatory) is operating at the Waste Isolation
Pilot Plant (WIPP, 1585 m w.e.) since May 2011. The experiment consists 
of 175 kg of Xe enriched to 80.6\% in $^{136}$Xe housed in a liquid time 
projection chamber (TPC).  Both ionization and 
scintillation are used to measure the energy with a resolution of 3.9 \% (FWHM) at
2,615 MeV. The detector is capable of effectively rejecting rays through 
topological cuts. 
%EXO--200 has recently claimed the first observation 
%of $2\nu\beta\beta$  in $^{136}$Xe (Q$_{\beta\beta}$ = 2458.7 keV) \cite{HAR12}. 
%Initial results on $0\nu\beta\beta$ decay together 
%with new result for 2$\nu$ mode are published in \cite{AUG12}.The fiducial volume 
%used in this analysis contains 79.4 kg of $^{136}$Xe (3.52$\cdot10^{26}$ atoms), 
%corresponding to 98.5 kg of active enrLXe. 
Results obtained after $\sim$ 3000 h of measurements are the following \cite{AUG12,ALB14}:
\vspace{0.5cm}

$T_{1/2}$ (2$\nu$, $^{136}$Xe) = $[2.165 \pm 0.016 (stat) \pm 0.059 (syst)]\cdot10^{21} yr$  (4)

$T_{1/2}$ (0$\nu$, $^{136}$Xe) $> 1.6\cdot10^{25} yr$ \footnote{Recently more weaker limit was 
published using $\sim$ 3 times higher statistics, 
$T_{1/2}$ (0$\nu$, $^{136}$Xe) $> 1.1\cdot10^{25} yr$ \cite{ALB14a}. }                                      (5)

\vspace{0.5cm}

%Last result provides upper limit $\langle m_{\nu}\rangle$ $<$ 0.14-0.38 eV 
%depending of NME values. 
With the present background, the predicted EXO--200 sensitivity
after 5 years of data taking will 
be $T_{1/2} \sim 4\cdot10^{25}$ yr ($\langle m_{\nu}\rangle$ $\sim$ 0.10--0.24 eV). 
%is project is also a prototype for a planned 1 tone sized experiment 
%that may include the
%ability to identify the daughter of $^{136}$Ba in real time, effectively 
%eliminating all classes of background except that due to 2$\nu$ decay 
%(see Section 3.5).

\subsection{KamLAND-Zen}

The detector KamLAND--Zen is a modification of the existing KamLAND 
detector carried out in the summer of 2011. The $\beta\beta$ source/detector is 13 tons 
of Xe-loaded liquid scintillator (Xe--LS) contained in a 3.08 m diameter 
spherical Inner Balloon (IB). The IB is suspended at the center of the KamLAND 
detector. 
%The IB is surrounded by 1 kton of liquid 
%scintillator (LS) contained in a 13 m diameter spherical Outer Balloon (OB) 
%made of 135 $\mu$m thick composite film. 
The outer LS acts as an active shield 
for external $\gamma$-rays and as a detector for internal radiation from the Xe--LS or IB. 
The Xe--LS contains (2.52 $\pm$ 0.07) $\%$ by weight of enriched xenon gas (full 
weight of xenon is $\sim$ 
330 kg). The isotopic abundances in the enriched xenon is
(90.93 $\pm$ 0.05) $\%$ of $^{136}$Xe. 
%Scintillation light 
%is detected by 1,325 17-inch and 554 20-inch photomultiplier tubes (PMTs). The energy resolution is 9.9 \% (FWHM)
%at 2.458 MeV.
%The energy spectrum of $\beta\beta$ decay candidates is shown 
%in Fig. 5. Unexpectedly detected background (BI $\approx$ 10$^{-4}$ counts/keV$\cdot$kg$\cdot$yr) 
%is $\sim$ two order of magnitude higher than estimated background using previous 
%data of KamLAND detector. 
%Nevertheless, the  2$\nu$ decay of 
%$^{136}$Xe has been measured corresponding to a  $^{136}$Xe $2\nu\beta\beta$  
%decay half-life of \cite{GAN12a} of :
Main obtained results are the following \cite{GAN12,GAN13}:

\vspace{0.5cm}
     
$T_{1/2}$ (2$\nu$, $^{136}$Xe) = $[2.30 \pm 0.02 (stat) \pm 0.12 (syst)]\cdot10^{21} yr$ (6)

$T_{1/2}$ (0$\nu$, $^{136}$Xe) $> 1.9\cdot10^{25} yr$ (7)

$T_{1/2}$ (0$\nu\chi^0$, $^{136}$Xe) $> 2.6\cdot10^{24} yr$ (8)  

\vspace{0.5cm}

%This is consistent with the result obtained by EXO--200. For $0\nu\beta\beta$ decay, 
%the data give a
%lower limit of $T_{1/2}$ (0$\nu$, $^{136}$Xe) $> 6.2\cdot10^{24}$ yr (90$\%$ C.L.) \cite{GAN12a}, 
%which corresponds 
%to limit, $\langle m_{\nu}\rangle$ $< 0.22-0.6$ eV. 

Now the Collaboration undertakes efforts to understand and decrease the background. 
In principle, it could be lowered by factor $\sim$ 100 (in this case sensitivity will be 
improved by factor $\sim$ 10). 
%If it will be done, sensitivity of experiment will essentially increase 
%and for 3 years of measurements will be $T_{1/2} \sim 2\cdot10^{26}$ yr that corresponds 
%to a sensitivity to the neutrino mass, $\langle m_{\nu}\rangle$ $\sim 0.04 - 0.11$ eV. After the end of the 
%first phase of the experiment a second phase is planned (see the section 3.6).

\begin{table}[h]
%\setcaptionmargin{0mm} \onelinecaptionsfalse
%\captionstyle{flushleft} 
\caption{Seven most developed and promising projects. 
Sensitivity at 90\% C.L. for three (1-st step of GERDA and MAJORANA, 
first step of SuperNEMO, SNO+, CUORE-0 and KamLAND-Zen) 
five (EXO, SuperNEMO and CUORE) and ten (full-scale GERDA and MAJORANA) 
years of measurements is presented. M - mass of isotopes.}
%\vspace{0.5cm}
%\rule[-2mm]{0mm}{5mm}
%\begin{center}
\begin{tabular}{llllll}
\hline
Experiment & Isotope & M, kg & Sensitivity & Sensitivity & Status \\
& &  & $T_{1/2}$, yr & $\langle m_{\nu} \rangle$, meV &  \\
\hline
CUORE \cite{ART14,AGU14}  & $^{130}$Te & 11 & $5\times10^{24}$ & 230--570 & in progress \\ 
  &  & 200 & $10^{26}$ & 50--130 & in progress \\ 
GERDA \cite{ACK13} & $^{76}$Ge & 40 & $2\times10^{26}$ & 80--190 & in progress \\
& & 1000 & $6\times10^{27}$ & 15--35 & R\&D\\ 
MAJORANA  & $^{76}$Ge & 30 & $1.5\times10^{26}$ & 90--200 & in progress \\
\cite{ABR14} & & 1000 & $6\times10^{27}$ & 15--35 & R\&D \\ 
EXO \cite{GRA13} & $^{136}$Xe & 200 & $4\times10^{25}$ & 100--240 & in progress \\
& & 5000 & $2\times10^{27}$ & 14--33 & R\&D \\ 
SuperNEMO & $^{82}$Se & 7 & 6.5$\times10^{24}$ & 240--560 & in progress \\
 \cite{BAR12} &  & 100--200 & (1--2)$\times10^{26}$ & 44--140 & R\&D \\
KamLAND-Zen  & $^{136}$Xe & 320 & 2$\times10^{26}$ & 44--105 & in progress \\
 \cite{INO13}& & 1000 &  $6\times10^{26}$ & 25-60 & R\&D \\
SNO+ \cite{LOZ14} & $^{130}$Te & 800 & $10^{26}$ & 50--130 & in progress \\
 & & 8000 & $10^{27}$ & 16-40 &  R\&D \\
\hline
\end{tabular}
%\end{center}
\end{table}

\subsection{GERDA-I}

GERDA is a low-background experiment which searches for the neutrinoless double beta
decay of $^{76}$Ge, using an array of high-purity germanium (HPGe) detectors isotopically
enriched in $^{76}$Ge \cite{ACK13}. The detectors are operated naked in ultra radio-pure liquid
argon, which acts as the cooling medium and as a passive shielding against the external
radiation. The experiment is located in the underground Laboratori Nazionali
del Gran Sasso of the INFN (Italy, 3500 m w.e.). The Phase I of GERDA use eight
enriched coaxial detectors refurbished from Heidelberg-Moscow and IGEX experiments
 (corresponding to approximately 18 kg of $^{76}$Ge). The energy resolution is $\approx$ 4.5 KeV at 2.039 MeV. 
GERDA-I measurements have been started in November 2011 and stoped in May of 2013. Results of the measurement  
are \cite{AGO13,AGO13b}:
\vspace{0.5cm}

$T_{1/2}$ (2$\nu$, $^{76}$Ge) = $[1.84 \pm 0.14 (stat) \pm 0.10 (syst)]\cdot10^{21} yr$  (9)

$T_{1/2}$ (0$\nu$, $^{76}$Ge) $> 2.1\cdot10^{25} yr$ (10)                                     

\vspace{0.5cm}
 
%presented 
%in Fig. 6 (6.1 kg$\cdot$yr of data). 

%The 2$\nu$ decay signal of $^{76}$Ge is clearly visible with a half-life $T_{1/2}$
%  $\approx 1.88\cdot10^{21}$ yr, preliminary result). Background index in 0$\nu$ region is 
%$\sim$ 0.02 c/keV$\cdot$kg$\cdot$yr.
%Blind analysis will be applied to the 0$\nu$ region (which is closed now). First result 
%will be reported in the beginning of 2013. Expected Sensitivity of GERDA-I with present background 
%is $\sim 2\cdot10^{25}$ yr for one year of measurement. 
In 2014 new $\sim$ 20 kg of HPGe crystals 
will be added and experiment will be transformed to Phase II (GERDA-II). 
%Description of full scale GERDA experiment is done in Sec. 3.2.

\section{Future large-scale experiments}

Seven of the most developed and 
promising experiments which can
be realized within the next few years are presented in  
Table 4. The
estimation of the sensitivity to $\langle m_{\nu} \rangle$ is made using NMEs 
from \cite{SUH12,SIM13,BAR09,RAT10,ROD10,MEN09} and phase-space factor values 
from \cite{KOT12}.

%\section{Conclusion}

%\
%\section*{Acknowledgments}

%In this place, the authors are kindly asked to put their own
%acknowledgments in the case of need. Note that there are no section
%numbers for the Acknowledgments or References.

%%%%%%%%%%%%%%%%%%%%%%%%%%%%%%%%%%%%%%%%%%%%%%%%%%%%%%%%%%%%%%%%%%
% References
%%%%%%%%%%%%%%%%%%%%%%%%%%%%%%%%%%%%%%%%%%%%%%%%%%%%%%%%%%%%%%%%%%

\end{document}